\documentclass[preprint,12pt,square,numbers,sort&compress]{elsarticle}
\usepackage{usebib}

\usepackage[utf8]{inputenc}
\usepackage{cmap}
\usepackage[T1]{fontenc}
\usepackage[english]{babel}
\usepackage{indentfirst}
\usepackage[top=2.94cm,bottom=2.94cm,right=2.94cm,left=2.94cm]{geometry}
\usepackage{setspace}
\setcitestyle{citesep={,}}

\usepackage{etoolbox}  
\usepackage{fancyhdr}
\usepackage{footnote}
\usepackage{footmisc}
\usepackage{caption}
\DeclareCaptionFormat{myformat}{\fontsize{9}{10}\selectfont#1#2#3}
\usepackage[labelformat=simple]{subcaption}

\usepackage{url}
\usepackage{hyperref}

\usepackage{amsmath}
\usepackage{amsfonts}
\usepackage{amssymb}
\usepackage{amsthm}
\usepackage{textcomp}
\usepackage{amssymb}
\usepackage{float}
\usepackage{comment}

\usepackage{multicol}
\usepackage{multirow}
\usepackage{array}
\usepackage{booktabs}
\usepackage{tabularx}

\usepackage{listings} 
\usepackage{yfonts}
\usepackage[usenames]{color} 
\usepackage{enumitem}
\usepackage{fancyhdr}
\usepackage{comment}

\usepackage[usenames,dvipsnames,svgnames,table]{xcolor}

\usepackage{pdfpages}
\usepackage{epsfig}
\usepackage{graphicx}
\usepackage{tikz}
\usetikzlibrary{fit}
\usepackage{wrapfig}
\usepackage{epstopdf}
\usepackage{subcaption}

\usepackage{algorithm}
\usepackage{algpseudocode}

\usepackage{makeidx}
\usepackage[title]{appendix} 
\makeindex

\usepackage{lmodern} \normalfont
\DeclareFontShape{T1}{lmr}{bx}{sc} { <-> ssub * cmr/bx/sc }{}
\usepackage{mathrsfs}

\usepackage[all]{xy}

\begin{document}
	\begin{frontmatter}
		\title{Phase field modelling and simulation of damage occurring in human vertebra after screws fixation procedure}
		\author[1]{Deison Preve}
		\author[2]{Pietro Lenarda\corref{cor}}
		\author[3]{Daniele Bianchi} 
		\author[3]{Alessio Gizzi} 
		
		\address[1]{}
		\address[2]{IMT School for Advanced Studies Lucca, Piazza San Francesco 19, 55100 Lucca, Italy}
		\address[3]{Department of Engineering, University of Rome Campus Bio-Medico, 00128 Rome, Italy}
		
		\cortext[cor]{Corresponding author. Tel: ,  e-mail: pietro.lenarda@imtlucca.it}
		
		\begin{abstract}
			
			The present endeavor numerically exploits the use of a phase-field model to simulate and investigate fracture patterns, deformation mechanisms, damage, and mechanical responses in a human vertebra after the incision of pedicle screws under compressive regimes. Moreover, the proposed phase field framework can elucidate scenarios where different damage patterns, such as crack nucleation sites and crack trajectories, play a role after the spine fusion procedure, considering several simulated physiological movements of the vertebral body. A convergence analysis has been conducted for the vertebra-screws model, considering several mesh refinements, which has demonstrated good agreement with the existing literature on this topic. Consequently, by assuming different angles for the insertion of the pedicle screws and taking into account a few vertebral motion loading regimes, a plethora of numerical results characterizing the damage occurring within the vertebral model has been derived. Overall, the phase field results may shed more light on the medical community, which will be useful in enhancing clinical interventions and reducing post-surgery bone failure and screw loosening. 
			
		\end{abstract}
		
	\end{frontmatter}

	\section{Introduction}
	\label{Introduction}
	
	Originated from mathematical techniques based on $\Gamma-$convergence 
	\cite{braides1998approximation,braides2002} and tailored for the approximation of free discontinuity problems \cite{ambrosio1990approximation,ambrosio1992approximation}, the phase field regularization of brittle fracture proposed by Francfort and Marigo \cite{francfort1998revisiting} has attracted a remarkable attention within  the computational fracture mechanics community over the last decade. As compared to other computational methods for  damage and fracture simulation in materials and components, such as for instance the Crack Band Model \cite{bazant1982crack}, the Smeared Crack Model \cite{bavzant1988nonlocal}, non-local and diffuse damage models
	\cite{bazant1994damage,gerard1998coupled}, or gradient damage models \cite{geers1998strain}, the phase field approach offers an elegant solution for problems involving linear elastic fracture mechanics. This solution is pursued through an energy minimization which, in the $\Gamma-$convergence limit, consistently reproduces the Griffith theory of fracture. The phase field approach to fracture, further analysed in \cite{bourdin2000numerical,bourdin2008variational}, has been applied in a considerable series of works proposing comparisons with other non-local damage models and discussing several detailed aspects regarding the finite element implementation 
	\cite{kuhn2008phase,hakim2009laws,amor2009regularized,nguyen2015phase,msekh2015abaqus,jodlbauer2020parallel,wu2017unified}. In this context, it is worth recalling the fundamental contribution by Miehe and co-workers \cite{miehe2010phase,miehe2010thermodynamically} that were the first to propose a robust finite element implementation of the phase field for brittle fracture, specialized to account for damage irreversibility and based on a suitable degradation mechanism able to simulate situations involving tensile stress states. 
	
	The state-of-the-art literature on phase field models clearly shows that the approach is mature for technical applications. In this regards, Tannè et al. \cite{tanne2018crack} have recently assessed the capabilities of the phase field approach (see in particular the AT1 and the AT2 models) to predict crack nucleation from V-notches and from points with stress concentrations.
	Phase field modeling has numerous applications, especially in the field of biomedical engineering, particularly in the study of bone fractures. In this regard, fractures in the vertebrae can be challenging to analyze and treat due to their complex shapes and locations  \cite{bianchi2022osteolytic}. However, phase field modeling can help by simulating the fracture process and shedding light on the underlying mechanisms involved. In particular, phase field modeling can be used to investigate the impact of vertebral fixation and screw placement on the healing of vertebral fractures. By incorporating the presence of screws and other fixation devices into the simulation, researchers can evaluate the mechanical stability of the fracture site and predict the potential for screw loosening or failure. Additionally, researchers can gain insights into the effects of different placement strategies on the fracture propagation process.
	
	
	Quantitative and qualitative finite element models are being developed to evaluate damage patterns and predict crack propagation in biological tissues. So much so, recently, the capability of a plethora of phase field implementations for modelling and analysing crack growth in bone tissue has been successfully applied \cite{gustafsson2022phase}, thus phase field theory seems a promissory approach to assess bone fracture patterns. In this context, the biomechanical problem of characterizing damage and predicting crack trajectories in a human vertebra after the fixation procedure of pedicle screws (see Fig. \ref{fig:screwsandvertebra}) can be tackled utilizing a phase field method.
	
	Metallic pedicle screws are used in spine fusion procedures when the intervertebral discs were damaged by aging or any trauma, causing the vertebrae to rub against each other and compress the nerves that pass through them.Spinal fusion joins two or more diseased vertebrae together, preventing motion at the vertebral segment \cite{molinari2021biomechanical}. The screws are inserted into the respective pedicle regions, connecting the screws through a vertical rod, providing a means for gripping onto a vertebral segment and limiting its motion, resulting in stable spine fixation. Bone damage and screw loosening may occur due to various post-operative events \cite{ohba2019risk}. Furthermore, the mechanical behavior of the pedicle screw insertion angle has been experimentally investigated \cite{sterba2007biomechanical,inceouglu2011pedicle}. Nonetheless, finite element analysis allows for the assessment of numerous possible failure scenarios that may occur after the surgical procedure requiring screw insertion, as well as the characterization of fracture damage and estimation of mechanical responses under varying screws fixation angles \cite{molinari2021effect}. Thereby, the impact of screw configuration angular parameters in fracture pattern and stress distributions were investigated via computational model considering a stress-based criterion \cite{yosibash2010predicting,matsuura2014specimen}. In the study by \cite{bianchi2022osteolytic}, the mechanical behaviour of instrumented metastatic vertebra in presence of pedicle screws was investigated. The findings showed that the prediction of mechanical properties was affected by the size and location of the metastasis. It was also concluded that a metastasis situated near the screws produces a higher fracture load response compared to a metastasis far from the screws. As a consequence, a finite element phase field model is hereof implemented in order to validate numerically the existing outcomes in the literature resulting from the variation of the screws insertion angle in a human vertebra, with the aim of further developing a coupled model with two or more fused vertebrae.
	
The paper is organized as follows: Section \ref{sec:cadmodels} provides the CAD models of the screws and the vertebra, which are discretized and have their material properties set. It also covers the virtual insertion of pedicle screws into the vertebra. Section \ref{sec:Phase} introduces the phase field finite element method strategy implemented in this study. The staggered and monolithic phase field models are detailed in Section \ref{sec:FEschemes}. Section \ref{sec:modellingPF} presents the boundary conditions and parameter setup for the numerical phase field method. Additionally, it includes a mesh sensitivity analysis of the phase field framework applied to several discretized models of vertebra-screws. Finally, Section \ref{sec:computationalanalysis} presents the main results in terms of damage patterns. It considers various screws insertion trajectories for different vertebral movements and includes a discussion comparing the outcomes with other relevant works in this field.
	
	\section{Human vertebra-pedicle screws finite element model}
	\label{sec:cadmodels}
	
	\subsection{Models and materials}
	
\begin{figure}[H]\centering
		\subcaptionbox{Identification of CC and ML directions of the pedicle screws virtually inserted.\label{fig:screwsandvertebra}}
		{ \includegraphics[width=1\linewidth]{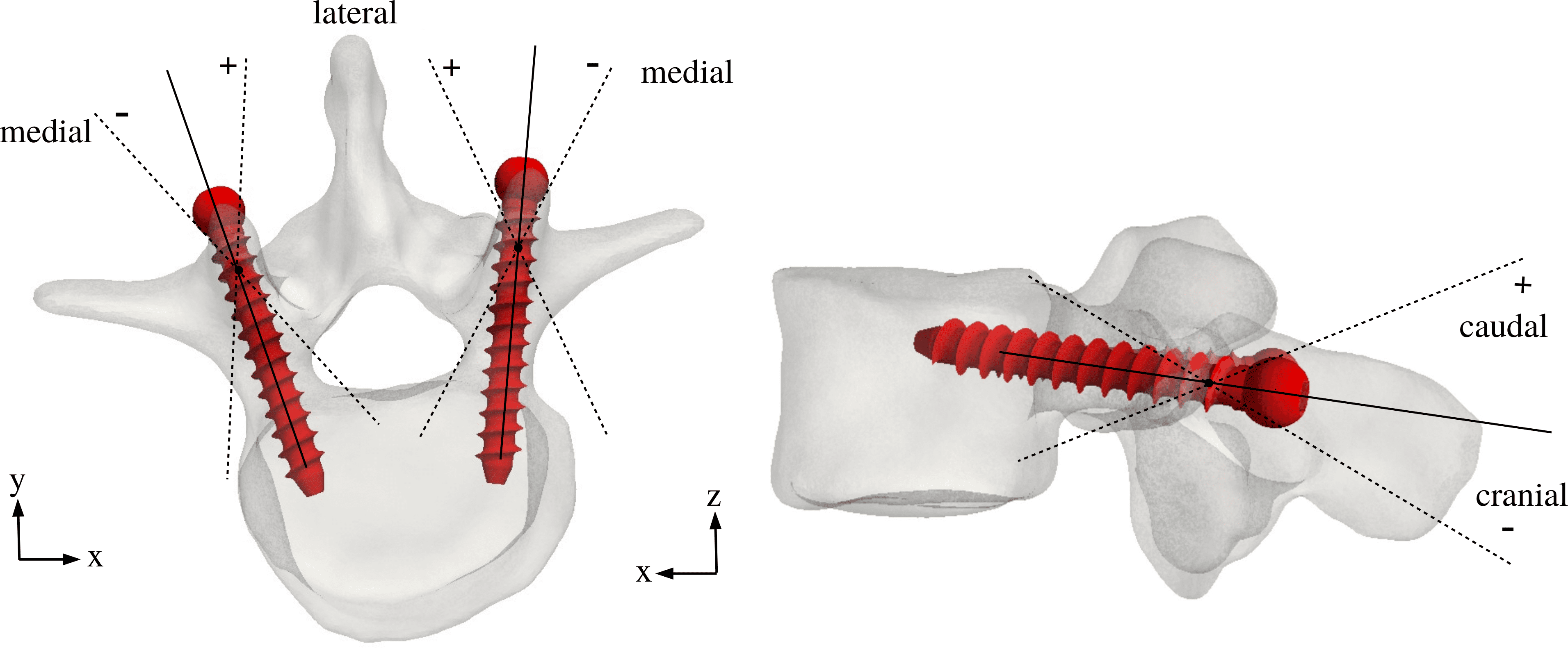}}
		\subcaptionbox{Uniform computational discretized vertebra-screws meshes. From left to right: 100 Dof, 350 Dof, 700 Dof. \label{fig:meshes}}
		{  	\includegraphics[width=1.0\linewidth]{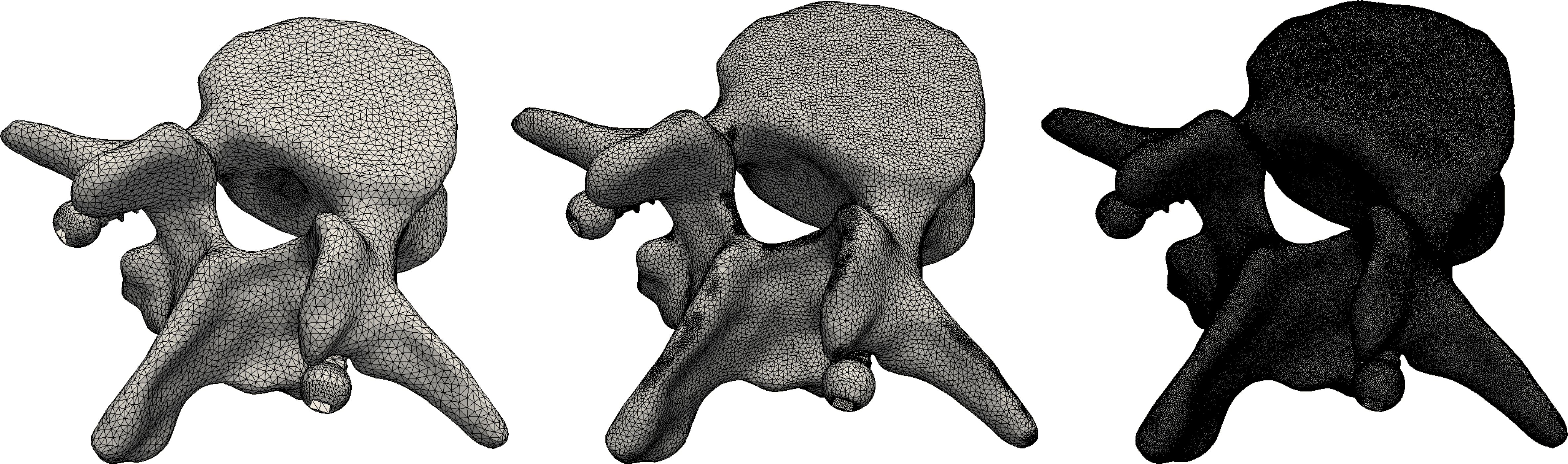}}
		\caption{CAD-based finite element of the vertebra-screws model.}
		\label{fig:CAD-FE-model}
	\end{figure}

	The L4 lumbar vertebra model was obtained from a Computer Tomography (CT) scan images from the spine of 49-year-old female patient, as described in \cite{molinari2021biomechanical}. The solid cylindrical pedicle screws have length of $40$ mm. The major and minor diameters of screws are $6.5$ mm and $4.3$ mm, respectively (please see \cite{molinari2021biomechanical, molinari2021effect} for additional details). The CAD models of the pedicle screws virtually inserted in the L4 vertebra are depicted in Fig. \ref{fig:CAD-FE-model}. The phase field model has been evaluated for three different screws fixation paths,  indicated by the vector ${\vec{\alpha}}=(\alpha_1,\alpha_2)$. Here, $\alpha_1$ represents the insertion angle in the cranio-caudal (CC) direction, and $\alpha_2$ indicates the insertion angle in the medio-lateral (ML) direction \cite{molinari2021biomechanical}, as shown in Fig. \ref{fig:screwsandvertebra}. In particular, the mesh sensitivity analysis is conducted using the screws insertion trajectory combination ${\vec{\alpha}}=(-5,0)$, which means that the screws are fixed with a negative angle of $-5^{\circ}$ in the CC direction and a neutral angle of $0^{\circ}$ in the ML direction.
	
	To import the combined vertebra-screws CAD model into the finite element environment \texttt{FEniCS} \cite{alnaes2011fenics}, uniform tetrahedral meshes are constructed. This is achieved by transferring the assembled STL model's triangular mesh into the pre-processing software HyperMesh \cite{altair2014hypermesh}, which generates the required tetrahedral solid. The solid is then assessed and converted into a MSH file using the Gmsh platform \cite{geuzaine2009gmsh}. Multiple refined meshes have been created, ranging from $60$ thousand elements to $700$ thousand elements. Please refer to Fig. \ref{fig:meshes} for visualization of these meshes.
	
\begin{figure}[H]\centering
		\subcaptionbox{Young’s modulus (GPa) distribution of the vertebra. The red color represents the high stiffness layer given by the cortical bone, while the blue color indicates low stiffness layer given by the trabecular bone.\label{fig:E1}}
		{  \includegraphics[width=.47\linewidth]{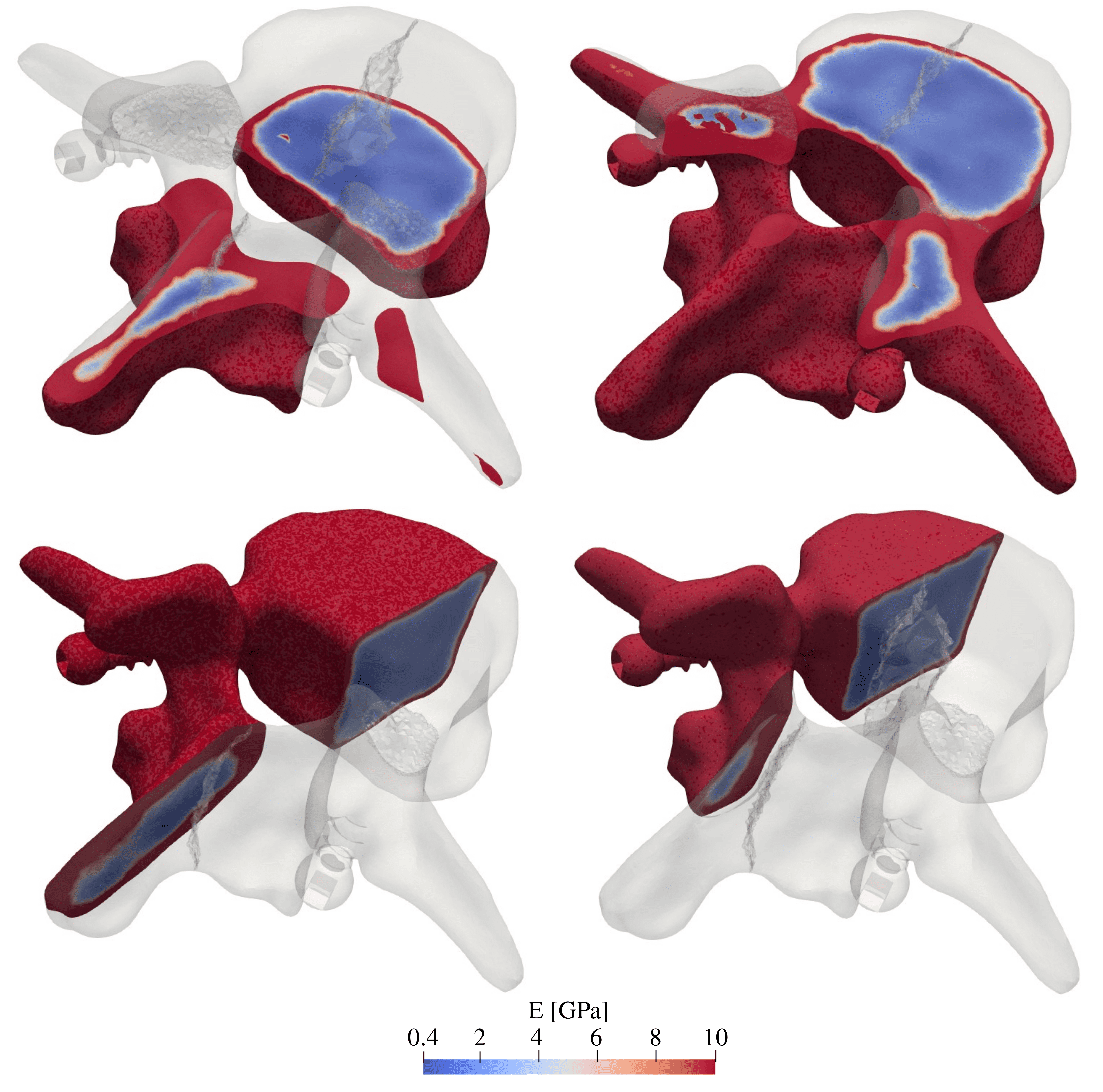}}
		\subcaptionbox{Young’s modulus (GPa) distribution of the pedicle screws. The red color illustrates the high stiffness assumed for the pedicle screws.\label{fig:E2}}
		{  \includegraphics[width=.5\linewidth]{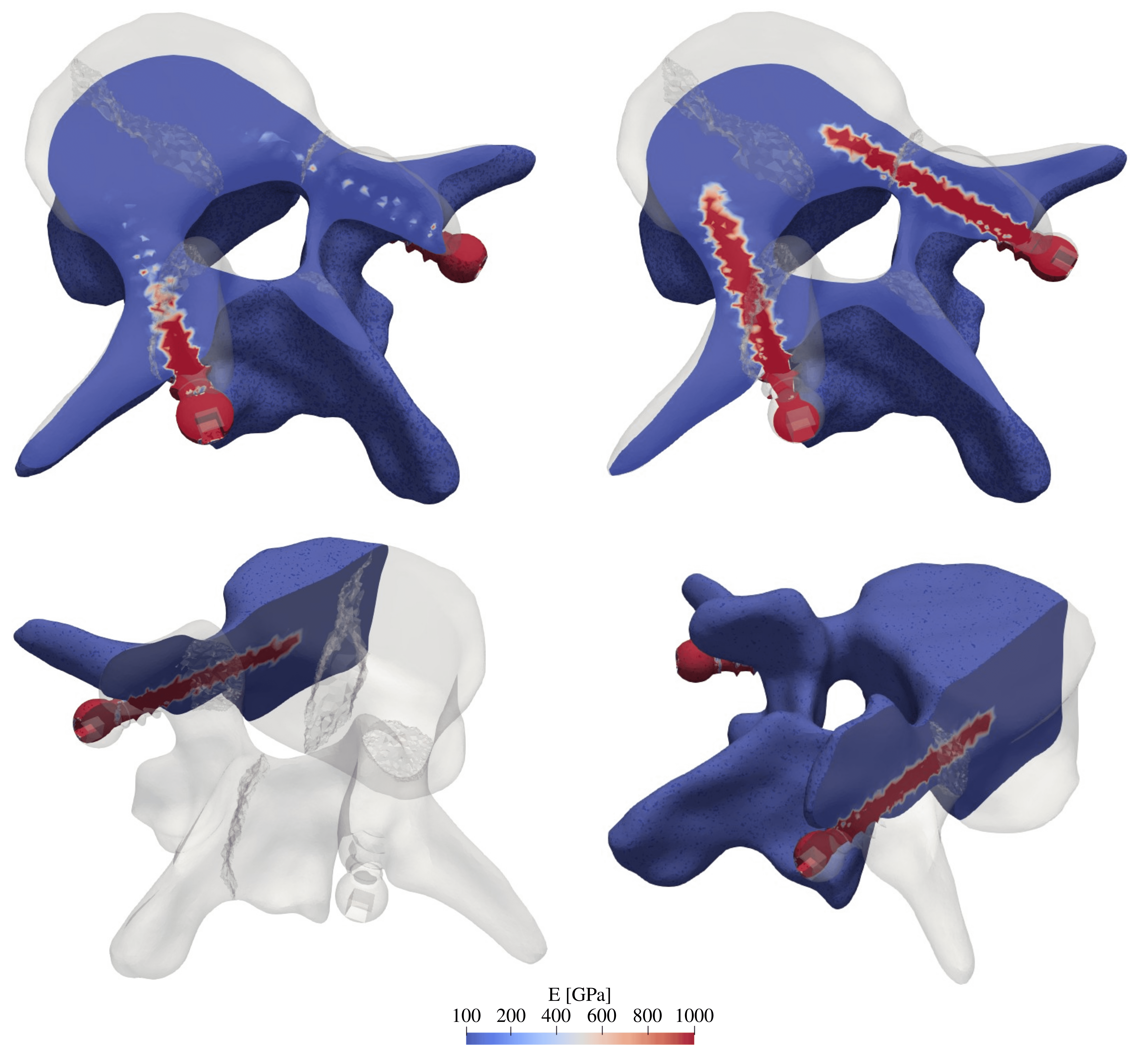}}
		\caption{Interpolated Young’s modulus (GPa) distribution of the model vertebra-screws. 
			correct stiffness values.}
		\label{fig:youngsmodulusdistribution}
	\end{figure}

	Bone density properties have been considered for the material of the L4 vertebra using CT data \cite{nalla2004effect,fields2010mechanisms,erdem2013simulation}. In this regard, a constant Poisson’s ratio of $\nu = 0.3$ was assumed, and isotropic and linear elastic material properties were adopted with a heterogeneous distribution of the Young’s modulus $E$. This distribution differentiates the cortical bone from the trabecular bone. For more details, please refer to \cite{molinari2021biomechanical,molinari2021effect}. For trabecular bone, an elastic modulus ranging between $0$ and $3$ GPa was derived, while for cortical bone, the range was between $12$ and $14$ GPa, as shown in Fig. \ref{fig:E1}. In what regards to the pedicle screws, surgical procedures commonly use biomedical implants or replacements made of the titanium alloy Ti-6Al-4V \cite{lin2014enhanced,song2021porous,aufa2022recent}. Therefore, the chosen mechanical properties for the pedicle screws are a Young's modulus of $110$ GPa and a Poisson's ratio of $0.4$.
	
	\section{Phase-field approaches to fracture with spectral decomposition}
	\label{sec:Phase}
	
	\noindent With reference to an arbitrary body occupying a domain $\Omega \in\mathbb{R}^{n_{dim}}$, with boundary $\partial\Omega\in\mathbb{R}^{n_{dim}-1}$, in the
	Euclidean space of dimension $n_{dim}$, in which an evolving internal discontinuity $\Gamma$ is postulated to exist, a material point is denoted by $\mathbf{x}$ and body forces  
	by $\mathbf{b}: \Omega \rightarrow\mathbb{R}^{n_{dim}}$.  Mixed  conditions are prescribed along non-overlapping Neumann and Dirichlet regions of the boundary  $\partial\Omega_{N}\cup\partial\Omega_{D}=\partial\Omega$ in the usual form
	\begin{equation}
		\label{BVP}
		\mathbf{u} = \overline{ \mathbf{u} } \hspace{0.2cm} \text{ on } \partial \Omega_{D}  \hspace{0.2cm} \text{ and } \hspace{0.2cm}     \boldsymbol \sigma \cdot \mathbf{n} = \mathbf{\overline{T}}  \hspace{0.2cm} \text{ on } \partial \Omega_{N} \ ,
	\end{equation}
	where $\mathbf{n}$ denotes the outward unit normal to the boundary, $\mathbf{u}$ is the displacement field and 
	$\boldsymbol \sigma$ is the Cauchy stress tensor, while $\overline{ \mathbf{u} }$ and $\mathbf{\overline{T}}$ are prescribed surface 
	displacements and tractions. 
	
	The variational approach to brittle fracture, governing crack nucleation, propagation and branching, is set up through the definition of the free energy functional \cite{miehe2010phase},
	
	\begin{equation}
		\Pi (\mathbf{u}, \Gamma) =  \Pi_{\Omega} (\mathbf{u}, \Gamma) + \Pi_{\Gamma} (\Gamma) ,
		\label{functional}
	\end{equation}
	embodying an additive decomposition between the elastic bulk energy $\Pi_{\Omega}$ stored in the damaged body and the energy $\Pi_{\Gamma}$ 
	necessary to nucleate and propagate a Griffith crack \cite{griffith1921vi}, defined as 
	
	\begin{equation}
		\Pi_{\Omega} (\mathbf{u}, \Gamma) + \Pi_{\Gamma} (\Gamma) =   \int_{\Omega \backslash \Gamma } \psi^{e}(\boldsymbol \varepsilon)  \ \mathrm{d}  \mathbf{x} + \int_\Gamma \mathcal{G}_{c}(\mathbf{u},s)\mathrm{d}\mathbf{x}\ ,
		\label{notregular}
	\end{equation}
	where $\psi^{e}$ is the elastic strain energy density, function of strain $\boldsymbol \varepsilon$, and $\mathcal{G}_{c}$ is the fracture energy, function of the displacement $\mathbf{u}$ and of the phase field variable $s$. 
	The latter parameter $s \in [0,1]$ is an internal state variable, ranging between 0 and 1 and representing isotropic damage, so that $s=0$ is representative of the intact material, while $s=1$ characterizes the fully damaged state.
	
	\subsection{The regularized variational formulation}
	\label{PFbulk}
	
	\noindent Within the regularized framework of the phase field approach \cite{bourdin2008variational,miehe2010phase, del2007variational,arroyo2006local},
	the potential energy of the system can be decomposed into two terms as following
	
	\begin{equation}\label{regular}
		\Pi (\mathbf{u}, s) = \int_{\Omega} \psi^e(\boldsymbol \varepsilon,s) \ \mathrm{d}  \mathbf{x}+\mathcal{G}_{c}\int_\Omega
		\gamma (s, \nabla s) \ \mathrm{d}  \mathbf{x} \ ,
	\end{equation}
	where $\psi^{e}(\boldsymbol \varepsilon,s)$ is the energy density of the bulk, now function of the damaged parameter $s$, and  $\gamma(s, \nabla s)$ is the crack density functional, with $\nabla $ denoting the spatial gradient operator. As a result, the total free energy density of the bulk $\hat{\psi}$ reads as
	\begin{equation*}
		\hat{\psi} (\boldsymbol \varepsilon,s)  =  \psi^{e}(\boldsymbol \varepsilon,s) +  \mathcal{G}_{c} \gamma(s, \nabla s).
	\end{equation*}
	The functional $\gamma (s, \nabla s)$ is assumed to be a convex function of $s$ and its gradient $\nabla s$ and can be written, in agreement with the following expressions characterizing the AT2 model,  respectively, as:
	\begin{equation}
		\gamma(s, \nabla s)=
		\dfrac{1}{2}\left(\dfrac{s^{2}}{l} + l|\nabla s |^{2}\right),\label{crackdensity}
	\end{equation}
	where $l_0$ stands for a regularisation characteristic length that can be related to the Young's modulus, the fracture toughness, and the tensile strength of the material, as specified in Section \ref{sec:modellingPF}. 
	
	\noindent To avoid the development of damage in compression, so to allow fracture growth only under tensile stress states, the following \lq tensile/compressive' decomposition is herein assumed for the energy density in the bulk $\psi^{e}(\boldsymbol \varepsilon,s)$ \cite{miehe2010phase, hofacker2013phase,miehe2015phasea, miehe2015phaseb,nguyen2016phase}
	and included in the formulation:
	
	\begin{equation}\label{energydensity}
		\psi^{e}(\boldsymbol \varepsilon,s) = g(s) \psi^{e}_{+}(\boldsymbol \varepsilon) + \psi^{e}_{-}(\boldsymbol \varepsilon),
	\end{equation}
	where $g(s)$ is a damage function that is assumed in the simple form $g(s) = \left(  1 - s \right)^{2}+k$,
	where $k$ is a residual stiffness (introduced to avoid ill-conditioning) and 
	the positive and negative parts of the energy density are defined as 
	\begin{equation}
		\label{pere}
		\psi^{e}_{\pm}(\boldsymbol \varepsilon)= \dfrac{\lambda}{2} \left(\text{tr} \boldsymbol \varepsilon_{\pm }  \right)^2 + \mu  \text{tr} \left(\boldsymbol \varepsilon_{\pm}^{2} \right) ,
	\end{equation}
	where $\lambda$ and $\mu$ are the Lam\'e constants, 
	$\text{tr} (\cdot)$ denotes the trace operator and the positive and negative parts of the strain $\boldsymbol \varepsilon_{\pm }$ are defined as follows. With reference to the spectral 
	representation for the strain (with eigenvalues 
	$\epsilon_i$ and unit eigenvectors $\boldsymbol e_i$), 
	denoted as 
	\begin{equation}
		\boldsymbol \varepsilon = \sum_{i=1}^3 \epsilon_i \boldsymbol e_i \otimes \boldsymbol e_i, 
	\end{equation}
	the strain is additively decomposed 
	as  $\boldsymbol \varepsilon = \boldsymbol \varepsilon_{+} + \boldsymbol \varepsilon_{-}$,
	so that the tensile and compressive parts associated to $\boldsymbol{\varepsilon}$ are
	\begin{equation}
		\boldsymbol \varepsilon_+ = \sum_{i=1}^3 \langle \epsilon_i \rangle \boldsymbol e_i \otimes \boldsymbol e_i, ~~~~~~ \mbox{ and } ~~~~~~
		\boldsymbol \varepsilon_- = -\sum_{i=1}^3 \langle -\epsilon_i \rangle \boldsymbol e_i \otimes \boldsymbol e_i,
	\end{equation}
	respectively, where the Macaulay bracket operator is defined for every scalar $x$ as $\langle x  \rangle = (x + | x |)/2$. 
	
	\noindent A standard derivation \cite{coleman1974thermodynamics} leads Eq. \eqref{pere} to the Cauchy stress tensor from the strain energy density:
	\begin{equation}
		\label{stress}
		\boldsymbol \sigma  = g(s) \boldsymbol \sigma_{+} + \boldsymbol \sigma_{-}=  \{ \left(  1 - s \right)^{2} + k \} \left(  \lambda  \text{tr} \boldsymbol \varepsilon_{+}  \, \mathbf{I}  + 2 \mu   \boldsymbol \varepsilon_{+} \right) +  \lambda  \text{tr} \boldsymbol \varepsilon_{-}  \, \mathbf{I}  + 2 \mu   \boldsymbol \varepsilon_{-},
	\end{equation}
	where $\mathbf{I}$ denotes the second-order identity tensor.
	The thermodynamic consistency of the above constitutive theory, in agreement with the Clausius-Duhem 
	inequality, has been addressed in \cite{miehe2010phase}.

	\subsection{Weak form of the variational problem}
	\label{Variational}
	
	The weak form corresponding to the phase field model for brittle fracture can be derived following a standard Galerkin procedure. In particular, the weak form of the coupled displacement and phase field damage problem according to Eq. \eqref{regular} is:
	
	\begin{equation}
		\delta \Pi=
		\int_{\Omega  } \boldsymbol \sigma ( \mathbf{u}) : \boldsymbol \varepsilon( \mathbf{v})    \ \mathrm{d}  \mathbf{x}  -  \int_{\Omega  }  2 \psi^e_+ (\boldsymbol \varepsilon) (1- s) \phi    \ \mathrm{d}  \mathbf{x}  + 
		\int_{\Omega  } \mathcal{G}_{c} \ \Big\{ \frac{1}{l}  s  \phi +   l\nabla  s \cdot \nabla \phi   \Big\}  \ \mathrm{d}  \mathbf{x}+ \delta \Pi_{\text{ext}} \; ,
		\label{var1}
	\end{equation}
	where $\mathbf{v}$ is the vector of the displacement test functions defined on $\textbf{H}^{1}_0(\Omega)$, $\phi$ stands for the phase field test function defined on $\textrm{H}^{1}_0(\Omega)$. Eq. \eqref{var1} holds for every test functions $\mathbf{v}$ and $\phi$. The external contribution to the variation of the bulk functional in Eq.~\eqref{var1} is defined as follows:
	\begin{equation}
		\label{var2}
		\delta \Pi_{\text{ext}} (\mathbf{u}, \mathbf{v}) =  \int_{ \partial \Omega}   \mathbf{\overline{T}} \cdot \mathbf{v} \ \mathrm{d} \Gamma +
		\int_{\Omega}   \mathbf{b} \cdot \mathbf{v}   \ \mathrm{d}  \mathbf{x}.
	\end{equation}
	
	\section{Numerical FE schemes}
	\label{sec:FEschemes}
	
	\subsection{Staggered solution scheme}
	\label{sec:staggered}
	\noindent The mechanical problem can be stated as: given the prescribed loading condition $\mathbf{\overline{u}}_n$ and $\mathbf{\overline{T}}_n$ at step $n$, find $\mathbf{u} \in \mathbf{U}  = \left\{\mathbf{u}\, | \,   \mathbf{u} = \overline{\mathbf{u}}_n \text { on }  \partial \Omega_{u} , \mathbf{u} \in \mathbf{H}^{1}(\Omega) \right\}$ such that
	\begin{equation}
		\label{varu}
		\mathcal{E}_{\mathbf{u}}(\mathbf{u},s; \mathbf{v}):=\int_{\Omega  } \boldsymbol \sigma(\mathbf {u}) : \boldsymbol \varepsilon( \mathbf{v})    \ \mathrm{d}  \mathbf{x}  - \int_{ \partial \Omega}   \mathbf{\overline{T}}_n \cdot \mathbf{v} \ \mathrm{d} \Gamma -
		\int_{\Omega}   \mathbf{b} \cdot \mathbf{v} \ \mathrm{d}  \mathbf{x}=0, \ \forall \mathbf{v} \in \mathbf{H}^{1}_0(\Omega),
	\end{equation}
	while the phase field problem is formulated as: find $s \in S  = \left\{ s \, | \, s = 1 \text { on }  \Gamma , s \in \mathrm{H}^{1}(\Omega) \right\}$ such that $\forall \phi \in \mathrm{H}^1_0(\Omega)$:
	
	\begin{equation}
		\mathcal{E}_{s}(\mathbf{u},s; \phi):=\int_{\Omega  } \mathcal{G}_{c} l\nabla  s \cdot \nabla \phi  \  \mathrm{d}  \mathbf{x}  + \int_{\Omega} \left( \dfrac{\mathcal{G}_{c}}{l} + 2 H \right) s \phi \ \mathrm{d}  \mathbf{x}-\int_{\Omega} 2 H \phi \ \mathrm{d}  \mathbf{x}=0 \; ,  \label{varphi2}
	\end{equation}
	where $H( \boldsymbol \varepsilon)=\text{max}_{\tau \in [0,t]}\left\lbrace \psi^e_+(\boldsymbol \varepsilon(\tau)) \right\rbrace$ is the strain history function, accounting for the irreversibility of crack formation \cite{miehe2010phase,msekh2015abaqus}. 
	
	To solve the quasi-static evolution problems for brittle fracture, isoparametric
	linear triangular finite elements are used for the spatial discretization, and a staggered solution scheme is considered. Staggered schemes based on alternate minimization exploit the convexity of the energy functional with respect to each individual variable $\mathbf{u}$ and $s$ \cite{wambacq2021interior}. Here, an 
	{\it ad hoc} developed solver has been implemented in the software \texttt{FEniCS}, see Alg. \ref{alg:staggered} for the staggered algorithm description. A series of benchmark tests taken from \cite{ambati2015review,miehe2010phase} has been carried out to validate the methodology.

	\subsection{Newton-Raphson procedure}
	\label{sec:newtonraphson}
	
	\noindent  Even if the mechanical problem has been split into Eqs.~\eqref{varu} and \eqref{varphi2}, so that the phase field is reduced to a linear problem, nonlinearity still remains, because of the piece-wise linearity of the constitutive law, which includes a spectral decomposition of the strain. Therefore, a consistent linearization is required, so that  the linear form defined by the residual can be written as:
	
	\begin{align}
		\begin{split}
			F_{\mathbf{u}}(\mathbf{u}, s ;\mathbf{v})=&\int_{\Omega  } \left\lbrace ( (1-s)^2+k ) \boldsymbol \sigma_{+}(\mathbf {u}) : \boldsymbol \varepsilon( \mathbf{v}) + \boldsymbol \sigma_{-}(\mathbf {u}) : \boldsymbol \varepsilon( \mathbf{v})  \right\rbrace \ \mathrm{d}  \mathbf{x}+ \\  
			& \quad \quad \quad \quad \quad \quad \quad \quad \quad \quad \quad \quad \quad \quad - \int_{ \partial \Omega}   \mathbf{\overline{T}} \cdot \mathbf{v} \ \mathrm{d} \Gamma -
			\int_{\Omega}   \mathbf{b} \cdot \mathbf{v}   \ \mathrm{d}  \mathbf{x}. 
		\end{split}
	\end{align}
	
	Given $\mathbf{u}^k$ the current Newton-Raphson approximate solution at iteration $k$, the correction $\delta \mathbf{u}$ is therefore the solution of the following linear variational problem: find $\delta \mathbf{u} \in \mathbf{U}_0  = \left\{\mathbf{u}\, | \,   \mathbf{u} = \mathbf{0} \text { on }  \partial \Omega_{u} , \mathbf{u} \in \mathbf{H}^{1}(\Omega) \right\}$ such that
	$
	J_{\mathbf{u}}(\delta \mathbf{u}, \mathbf{u}^k,s ; \mathbf{v})=-F_{\mathbf{u}}(\mathbf{u}^k, s; \mathbf{v}), 
	$ $\ \forall \mathbf{v} \in \mathbf{H}^{1}_0(\Omega) \ ,$
	and then iterate as $\mathbf{u}^{k+1}=\mathbf{u}^k + \delta \mathbf{u}$.
	The Jacobian entering the formulation is 
	\begin{flalign}
		J_{\mathbf{u}}(\delta \mathbf{u},\mathbf{u},s ; \mathbf{v} )=\int_{\Omega  } \left\lbrace ( (1-s)^2+k ) \partial \boldsymbol \sigma_{+}(\delta \mathbf{u}, \mathbf {u}) : \boldsymbol \varepsilon( \mathbf{v}) + \partial \boldsymbol \sigma_{-}(\delta \mathbf {u}, \mathbf{u}) : \boldsymbol \varepsilon( \mathbf{v})  \right\rbrace \ \mathrm{d}  \mathbf{x}\ ;&&
	\end{flalign}
	for details about the terms $\partial \boldsymbol \sigma_{-}$, $\partial \boldsymbol \sigma_{+}$, please see \cite{jodlbauer2020parallel}.
	
	\begin{algorithm}
		\caption{Staggered iterative scheme for phase field fracture at a step $n \geq 1$}\label{alg:staggered}
		\begin{algorithmic}[1]
			\State  \textbf{Input:} Displacements and phase field $(\mathbf{u}_{n-1}, s_{n-1})$ and prescribed loads $(\mathbf{\overline{u}}_{n} , \mathbf{\overline{T}}_{n})$: 
			\State Initialize $(\mathbf{u}^0, s^0):=(\mathbf{u}_{n-1}, s_{n-1})$;
			\For{$k \geq 1$ \ \text{staggered iteration}}: 
			\State Given $s^{k-1}$, solve the mechanical problem (13): 
			$\mathcal{E}_{\mathbf{u}}(\mathbf{u}, s^{k-1}; \mathbf{v})=0$ for $\mathbf{u}$, set $\mathbf{u}:=\mathbf{u}^k$;
			\State Given $\mathbf{u}^k$, solve the phase field problem (14): 
			$\mathcal{E}_{s}(\mathbf{u}^k, s; \phi)=0$ for $s$, set $s:=s^k$;
			\If {$\text{max}\{ || \mathbf{u}^k-\mathbf{u}^{k-1} ||/||\mathbf{u}^k ||, | s^k-s^{k-1} |/|s^k| \} < \text{tol}$:}
			\State set $(\mathbf{u}^k,s^k):=(\mathbf{u}_n, s_n)$;
			\Else \ $k+1 \rightarrow k$.
			\EndIf
			\EndFor
			\State  \textbf{Output:} $(\mathbf{u}_n, s_n)$.
		\end{algorithmic}
	\end{algorithm}
	
	At this stage, it is fundamental to remark that, in order to predict crack trajectories in human vertebrae under tensile/compressive stress states, the phase field finite element method will be formulated. This formulation involves decomposing the strain energy density $ \psi^e(\mathbf{u},s) $ in Eq. \eqref{energydensity} based on spectral diagonalization, as described in \cite{miehe2010phase}, into active and passive parts. This decomposition allows for the application of material response degradation only in tension. The variational formulation is then implemented in the \texttt{FEniCS} environment, utilizing the MPI (message passing interface) parallelization library, which accelerates the computational time.
	
	\subsection{Monolithic solution scheme}
	\label{sec:monolithic}
	
	A monolithic solver has also been also implemented in \texttt{FEniCS}. The tangent operator of the non-linear variational functional,
	\begin{align}
		\begin{split}
			\mathcal{E}_{\mathbf{u},s}(\mathbf{u},s; \mathbf{v}, \phi):=& \int_{\Omega  } \boldsymbol \sigma ( \mathbf{u}) : \boldsymbol \varepsilon( \mathbf{v})    \ \mathrm{d}  \mathbf{x}  -  \int_{\Omega  }  2 H(\boldsymbol{\varepsilon}) (1- s) \phi    \ \mathrm{d}  \mathbf{x} +\\
			& \int_{\Omega  } \mathcal{G}_{c} \ \Big\{ \frac{1}{l}  s  \phi +   l\nabla  s \cdot \nabla \phi   \Big\}  \ \mathrm{d}  \mathbf{x}
			+ \int_{ \partial \Omega}   \mathbf{\overline{T}} \cdot \mathbf{v} \ \mathrm{d} \Gamma +
			\int_{\Omega}   \mathbf{b} \cdot \mathbf{v}   \ \mathrm{d}  \mathbf{x},
		\end{split}
	\end{align}
	
	is computed via the symbolic derivative \texttt{derivative}, the monolithic algorithm scheme can be visualized in Alg. \ref{alg:mono}.
	
	\begin{algorithm}
		\caption{Monolithic iterative scheme for phase field fracture at a step $n \geq 1$}\label{alg:mono}
		\begin{algorithmic}[1]
			\State  \textbf{Input:} Displacements and phase field $(\mathbf{u}_{n-1}, s_{n-1})$ and prescribed loads $(\mathbf{\overline{u}}_{n} , \mathbf{\overline{T}}_{n})$: 
			\State Solve the coupled non-linear variational problem via Newton-Raphson iterative scheme: 
			\State $ \mathcal{E}_{\mathbf{u}, s}(\mathbf{u}_n, s_n; \mathbf{v}, \phi)=0, \quad \forall (\mathbf{v}, \phi) \in \mathbf{H}^1_0(\Omega) \times \mathrm{H}^1_0(\Omega)$
			\State  \textbf{Output:} $(\mathbf{u}_n, s_n)$.
		\end{algorithmic}
	\end{algorithm}

	\section{Phase Field modelling to pedicle screws in human vertebra}
	\label{sec:modellingPF}
	
	\subsection{Parameters calibration}
	\label{sec:parameterscalibration}
	
	The length scale $l_0$ parameter is deeply inserted for modelling phase field, considering that for a sufficiently small length scale $ l ,$ the functional \eqref{var1} converges to the total potential energy functional  \eqref{functional}, in the sense that the global minimizers of $ \Pi_l $ will also converge to that of $ \Pi. $ This entails that the length scale must be carefully chosen, rather than setting it arbitrarily. Having said that, on one hand, the characteristic length scale  is related to the apparent material strength \cite{pham2011gradient,martinez2018phase,preve2022comprehensive}. Particularly, once the base material properties are attributed, namely Young's modulus $ E, $ critical energy release rate $ \mathcal{G}_{c},$ then the characteristic length  can be tuned as 
	
	\begin{equation}\label{lmaterial}
		l=\dfrac{27}{256}\left(\dfrac{\mathcal{G}_{c}E}{\sigma^2_{\max}}\right).
	\end{equation}
	Consequently, the failure stress $ \sigma_{\max} $ was considered according to \cite{keller1994predicting,keyak1994correlations,molinari2021biomechanical}, so the length scale  can be evaluated through Eq. \eqref{lmaterial}.
	
	
	On the other hand, the characteristic length scale  can be interpreted, and therefore calibrated, as a structural factor. In particular, in a biological material, as cortical bone for instance, the length scale  can be attributed to the lamellar microstructure, which may trap the crack tip within it, deflecting or stopping the leading edge of the crack \cite{zysset1999elastic}. Moreover, structural changes in cortical bone due to ageing affect crack path and damage properties \cite{nalla2004effect,gustafsson2019age,gustafsson2019crack}.
	
	In essence, the characteristic length scale  that smears the regularized crack can be obtained either from microscopic structural-related mechanisms \cite{chan2009relating,koester2011effect}, or from bone properties \cite{shen2019novel,hug2022predicting}. As such, prior to determine the characteristic length scale  through Eq. \eqref{lmaterial}, a power-law equality is usually assumed by which the critical energy release rate $\mathcal{G}_c$ is derived from bone density properties, namely
	
	\begin{equation}
		\label{gccalibration}
		\mathcal{G}_c=\mathcal{G}_{c_0}\left(\dfrac{E}{E_0}\right)^\beta,
	\end{equation}
	where the parameters have been set as $E_0=20.000 \; \rm MPa$ $\mathcal{G}_{c_0}=0.01 \; \rm N/mm, $ and $\beta=0.8 $ are the base elastic modulus, the base critical energy release rate, the bone ash density and the power-law exponent, respectively. For those reasons, when a mesh sensitivity analysis is being conducted in a biological material, the phase field parameters must also be carefully examined and validated. As such, it has been performed in \cite{hug2022predicting,gustafsson2022phase}, different values for the characteristic length scale  have been considered, where the latter has calibrated phase field parameters after mechanical in-vitro experiments on human humeri bones. In addition, the critical energy release rate $ \mathcal{G}_c $ plays a crucial role in phase field modelling for bone application, where throughout experiments, different values of $ \mathcal{G}_c $ have been obtained at different aged cortical bones \cite{gustafsson2019age,shen2019novel}. Following these aspects, the critical energy release rate $ \mathcal{G}_c $ and the length scale  have been calibrated using Eqs. \eqref{lmaterial} and \ref{gccalibration}, and validated from the experimental studies herein mentioned.
	
	\subsection{Boundary conditions, numerical implementation and mesh sensitivity analysis}
	\label{sec:boundaryconditionsmeshsensititvity}
	
\begin{figure}[H]\centering
		{ 	\includegraphics[width=.8\linewidth]{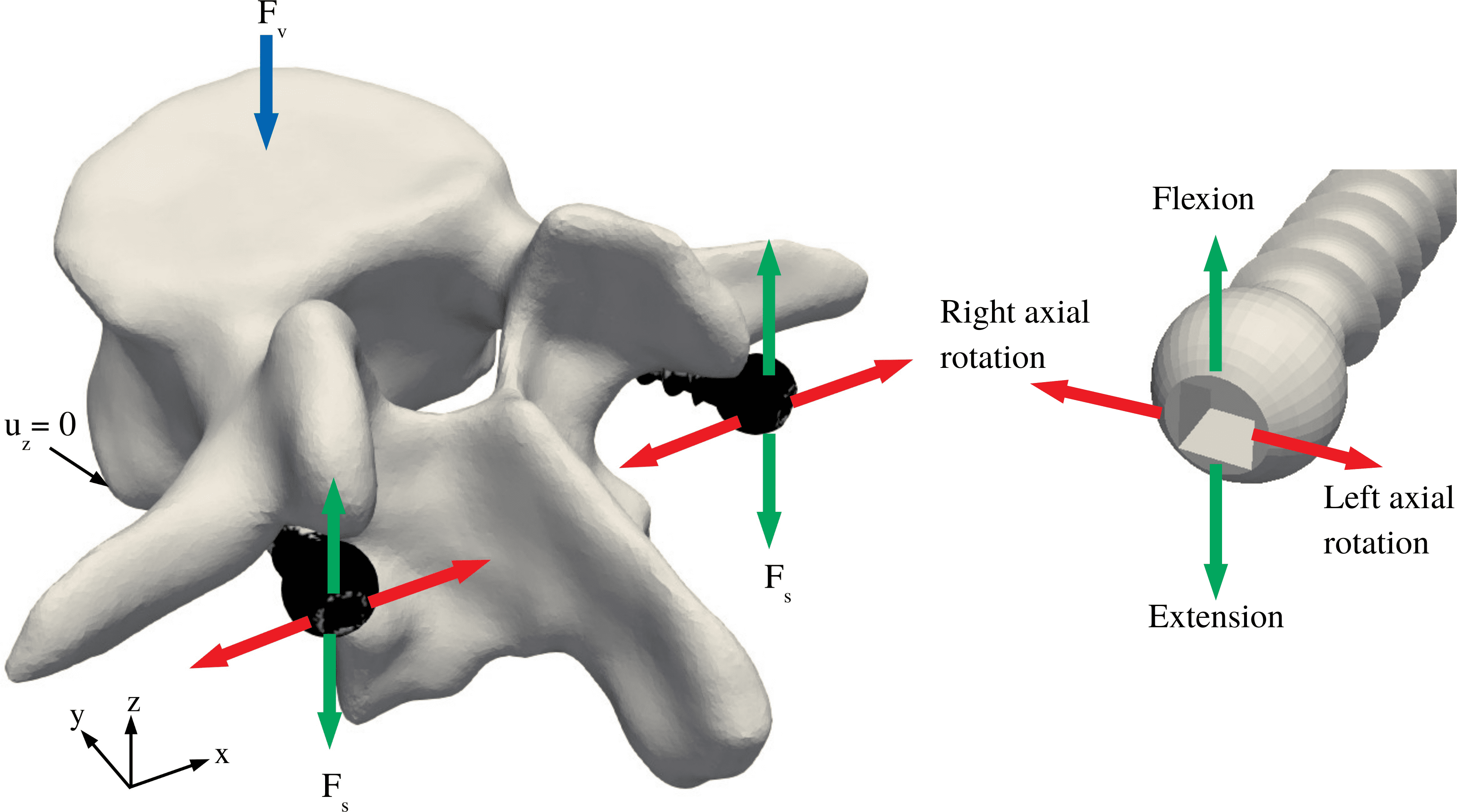}}
		\caption{Boundary conditions and loading regimes employed on the vertebra-screws models for numerical implementation.\label{fig:screwsandvertebra1}}
	\end{figure}

	Although spine models have been extensively exploited and validates in the literature \cite{dreischarf2014comparison}, the present finite element phase field method focuses on a single vertebral body, as observed in other studies  \cite{matsukawa2015biomechanical,molinari2021effect}. In this study, the loadings are addressed on the inserted screws in the L4 vertebra, which will be simultaneously constrained. Throughout all the numerical implementations, the boundary conditions were set to replicate the principal movements permitted by the vertebral column. For this purpose, the pedicle screws head have been loaded to reproduce flexion (bending forward) and extension (bending backwards) (green arrows in Fig. \ref{fig:screwsandvertebra1}), rotation (torsion/twisting) (red arrows in Fig. \ref{fig:screwsandvertebra1}). The centre of the L4 vertebra has been assumed as the rotation axis centre for the twisting loading mode. Meanwhile, a compressive force of $F_v=5$ N has been applied to the superior endplate of the vertebral body, while the inferior endplate has been constrained. Furthermore, the vertebral fracture patterns have been simulated by incrementally applying a quasi-static force to the screw heads for all the studied movements. Fundamentally, the total force applied at each loading step corresponds to $10\%$ of the constant compressive load $F_v.$
	
	The numerical phase field procedures were implemented in a High-Performance Computing (HPC) system, utilizing parallel computing and dividing the model into three smaller tasks. The computational time ranged from 2 hours for the coarsest mesh to 30 hours for the finest mesh. Six cores were utilized, with an average of 200 GB of RAM per core. It is worth mentioning that the extension and flexion loading modes required more time for simulation, necessitating the use of additional cores and computational memory to improve the efficiency of these movement modes.

\begin{figure}[H]\centering
		\subcaptionbox{Force vs. Displacement curves. \label{fig:meshsensitivity1}}
		{  \includegraphics[width=.45\linewidth]{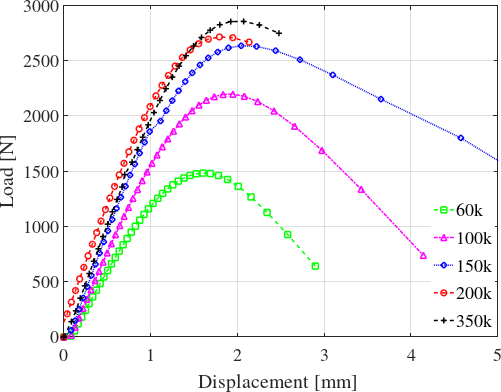}}
		\subcaptionbox{Relative error as function of the number of degrees of freedom. \label{fig:meshsensitivity2}}
		{ \includegraphics[width=.45\linewidth]{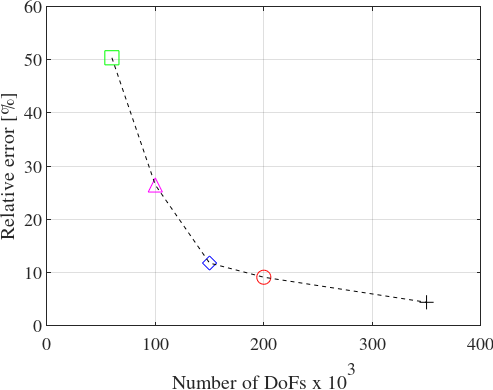}}
		\subcaptionbox{Vertebral fractured volume as function of the loading steps.\label{fig:meshsensitivity3}}
		{ \includegraphics[width=.45\linewidth]{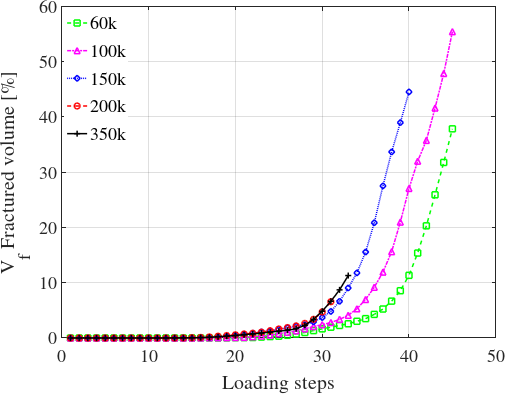}}
		\caption{Mesh sensitivity analysis: Flexion mode movement at screws fixation angle of $\vec\alpha=(-5,0).$}
		\label{fig:meshsensitivity}
	\end{figure}

	With the aim of achieving an appropriate mesh discretization for obtaining accurate numerical phase field responses for various physiological motions, a meticulous mesh sensitivity analysis has been conducted using the phase field finite element method. The vertebra-screws mesh model has been uniformly discretized at various refinements levels, ranging from $60.000$ to $350.000.$ Fig. \ref{fig:meshsensitivity}presents the numerical analysis comparing the outcomes in the flexion loading mode for different mesh refinements, with the screws insertion angle set as $\vec\alpha=(-5,0)$. Fig. \ref{fig:meshsensitivity1} depicts the Load vs. Displacement curves juxtaposed, indicating that as the mesh becomes finer, the vertebra-screws model can bear more load. Additionally, Fig. \ref{fig:meshsensitivity2} illustrates the relative error as the number of elements increases. According to \cite{molinari2021biomechanical}, it can be noticed that mesh convergence is satisfactorily achieved at a relative error of $5\%.$ Therefore, assuming a mesh size between $200.000$ and $350.00$ elements is acceptable for characterizing the fracture patterns in the simulated physiological movements. Nonetheless, in terms of characterizing the damage patterns, the present mesh sensitivity analysis revealed no significant distinctions among the different mesh refinements. Moreover, the computational cost aligns well with the numerical outcomes using a mesh with $200.000$ elements. This is further supported by Fig \ref{fig:meshsensitivity3} which displays the fractured volume as the loading step varies.
	These convergence analyses demonstrate good agreement with previous studies \cite{molinari2021biomechanical,molinari2021effect}. In terms of fracture type, although the phase field model captured similar damage patterns for all mesh refinements, coarser meshes exhibited more rapid fracture spread within the cortical part.
	
\begin{figure}\centering
		{\includegraphics[width=.45\linewidth]{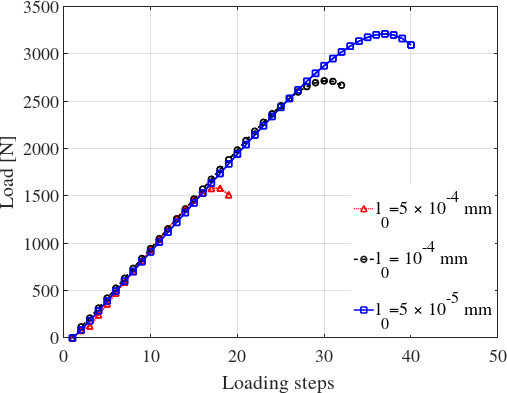}}	
		\caption{Load vs. loading steps obtained for different values of the length scale parameter $l_0$. Flexion mode movement at screws fixation angle of $\vec\alpha=(-5,0).$}
		\label{fig:meshsensitivity4}
	\end{figure}

	In accordance with the estimations made in \cite{hug2022predicting}, an examination of the role played by the characteristic length scale $l_0$ is also developed here. Taking into consideration the previous mesh sensitivity analysis, with a mesh of $200.000$ elements, a screws insertion angle of $\vec\alpha=(-5,0)$ in a flexion mode motion, Fig. \ref{fig:meshsensitivity4} indicates that the apparent peak load from the phase field scheme increases as the characteristic length scale $l_0$ is reduced, as expected based on previous results reported in the literature \cite{paggi2017revisiting,kumar2021phase}.	
	
	\section{Computational analysis}
	\label{sec:computationalanalysis}
	
	\subsection{Characterization of loading modes}
	\label{sec:characloadingmodes}
	
\begin{figure}[H]\centering
		\subcaptionbox{Force vs. Displacement curves at flexion mode motion. \label{fig:FvDflexion}}
		{ \includegraphics[width=.45\linewidth]{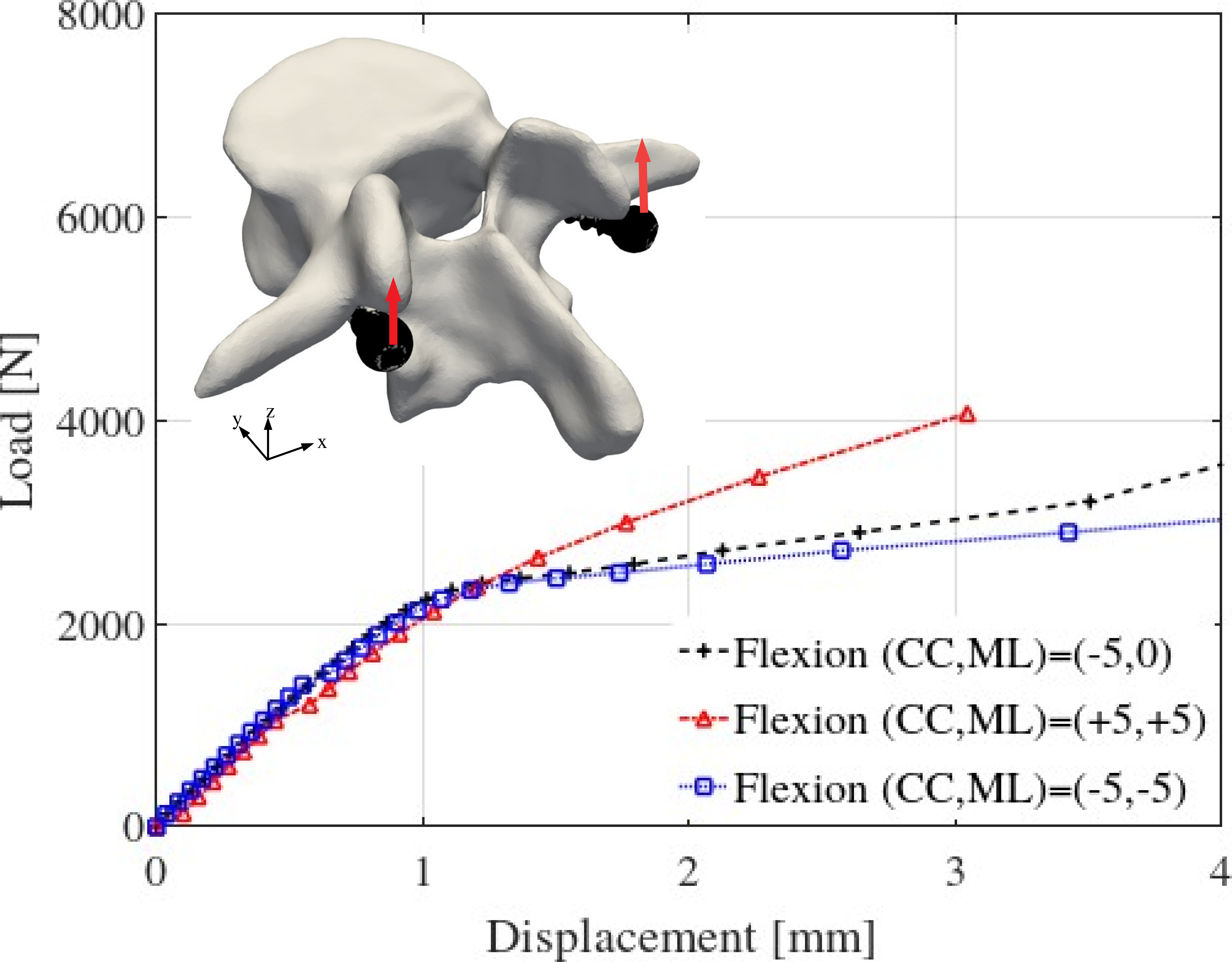}}
		\subcaptionbox{\label{fig:F1}}
		{  \includegraphics[width=.45\linewidth]{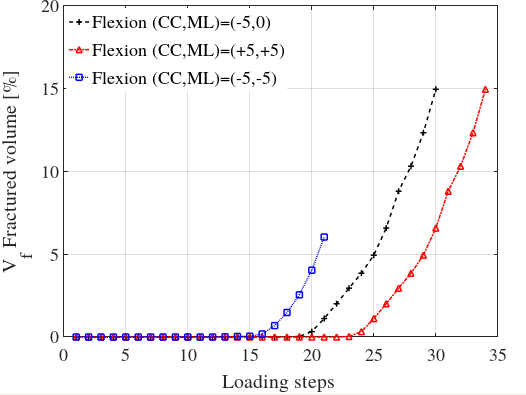}}
		\subcaptionbox{Force vs. Displacement curves at extension mode motion. \label{fig:FvDextension}}
		{ \includegraphics[width=.45\linewidth]{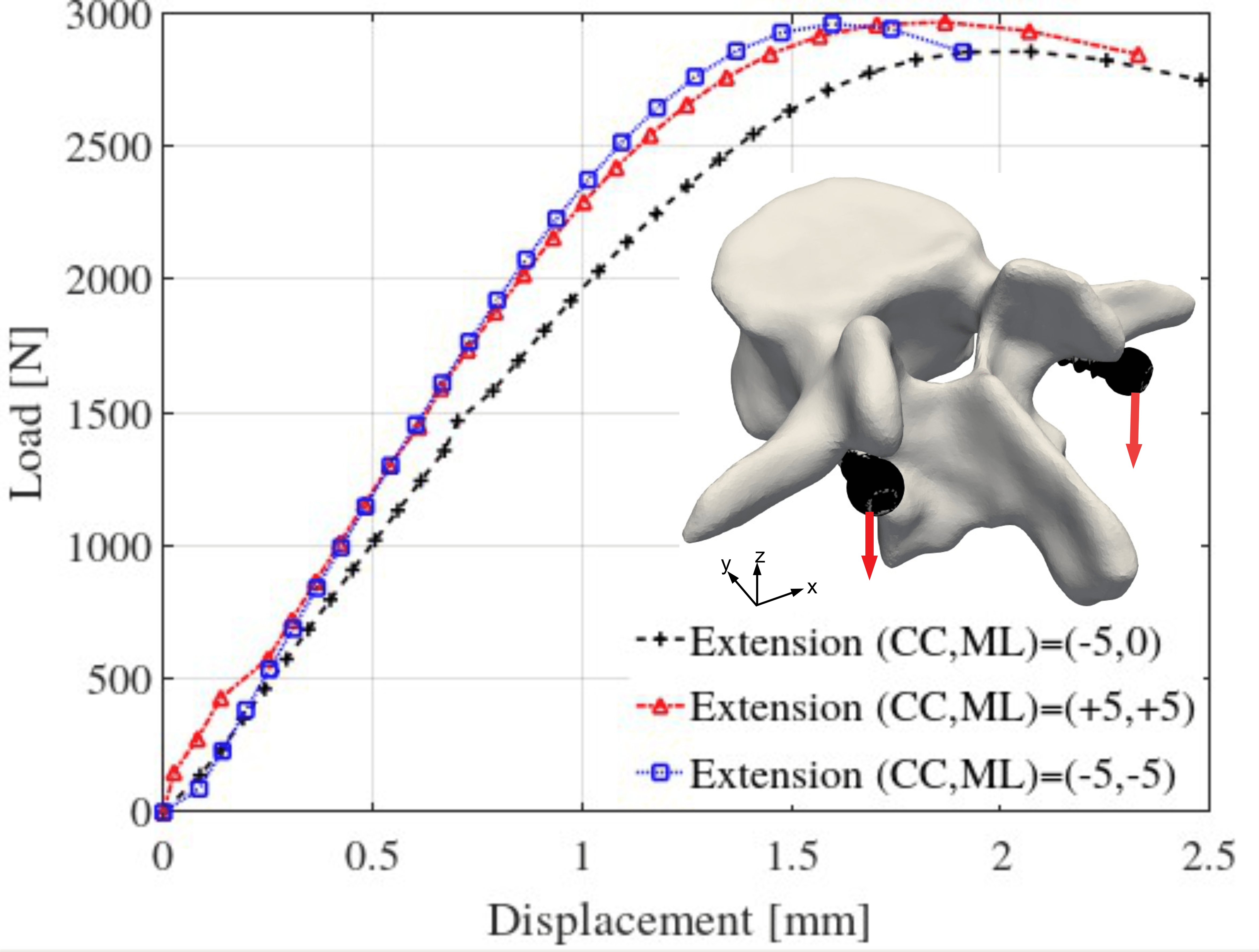}}
		\subcaptionbox{\label{fig:F3}}
		{ \includegraphics[width=.45\linewidth]{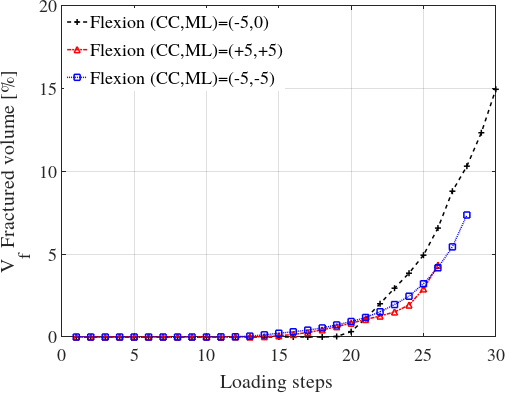}}
		\subcaptionbox{Force vs. Displacement curves at torsion mode motion.\label{fig:FvDtorsion}}
		{  \includegraphics[width=.45\linewidth]{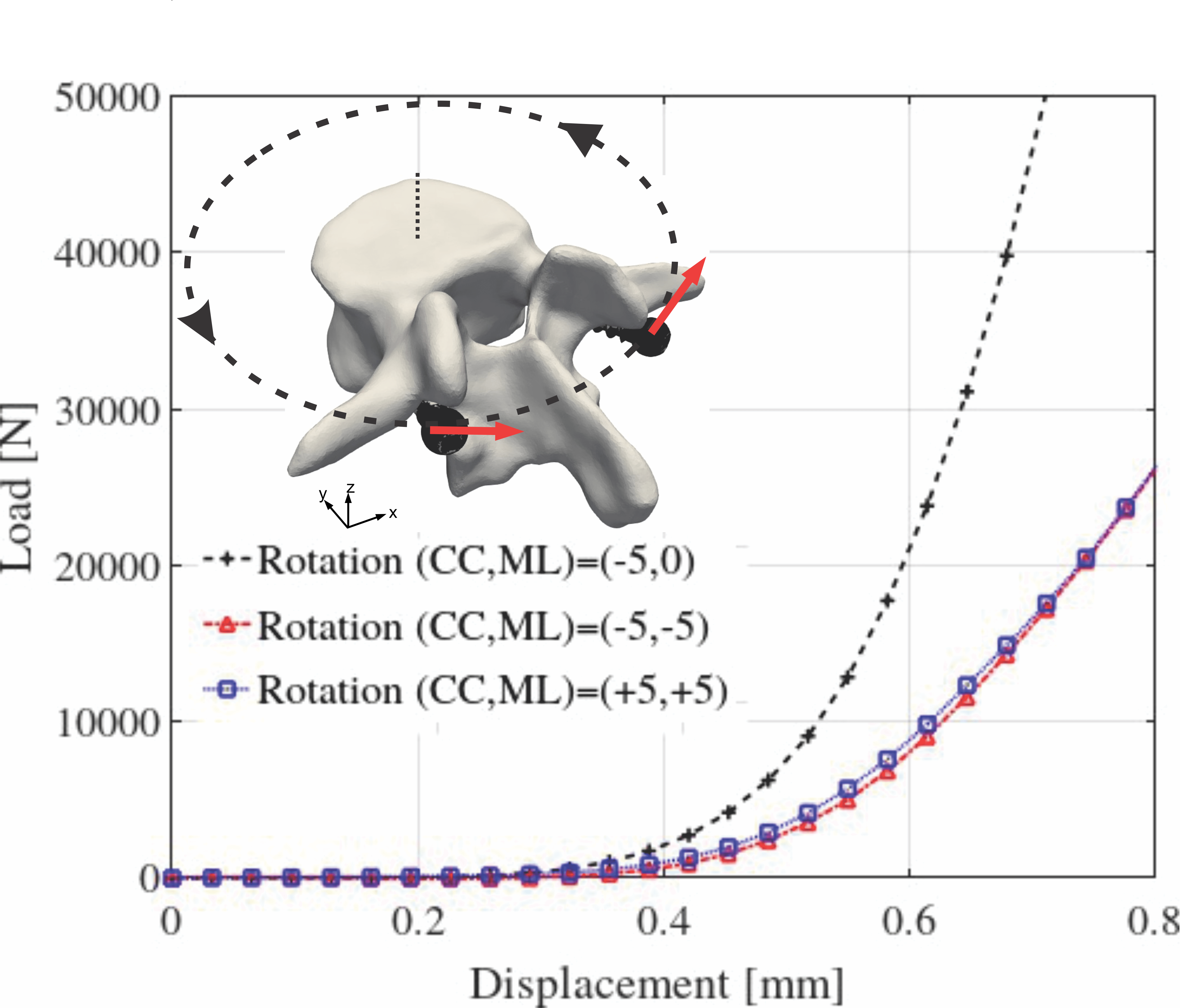}}
		\subcaptionbox{\label{fig:F5}}
		{ \includegraphics[width=.45\linewidth]{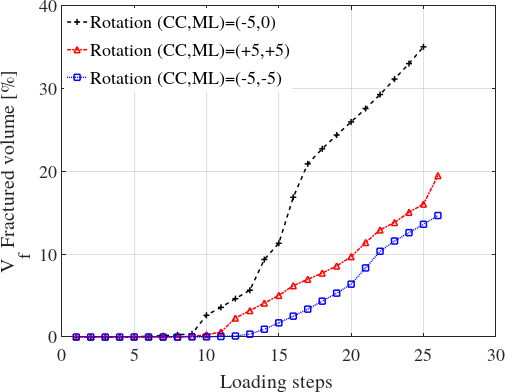}}
		\caption{Comparative analyses among the flexion, extension and torsion modes, at screws fixation angles of ${\vec{\alpha}}=(-5,0),$ ${\vec{\alpha}}=(+5,+5)$ and ${\vec{\alpha}}=(-5,-5)$.}
		\label{fig:Comparative}
	\end{figure}

	Continuing with the comparative investigation of the various screw insertion angles, namely ${\vec{\alpha}}=(-5,0),$ ${\vec{\alpha}}=(+5,+5),$ and ${\vec{\alpha}}=(-5,-5).$ Fig. \ref{fig:Comparative} highlights how the screw configuration influences the mechanical responses of the vertebral body in different motion regimes. The Force vs. Displacement curves generated from the phase field finite element method are displayed in Figs. \ref{fig:FvDflexion}, \ref{fig:FvDextension}, and \ref{fig:FvDtorsion}, representing the vertebral movements of flexion, extension, and torsion, respectively. 
	\\
	In the three cases analyzed, the Load is computed as the vincular reaction on the bottom of the vertebra. In the cases of flexion and extension there is a linear elastic phase, followed by a softening due to the occurrence of fracture inside the vertebra, whereas in the case of rotation, the vincular reaction on the bottom is equal to zero, until a critical angle is reached and the fracture is activated.
	
	
	Regarding the damage patterns, the counter-clockwise torsion motions simulated for the insertion angles ${\vec{\alpha}}=(-5,0),$ ${\vec{\alpha}}=(+5,+5),$ and ${\vec{\alpha}}=(-5,-5)$ were characterized by the formation of asymmetric damage, where damage occurrence is more prominent on the side that experiences greater loading as the steps increase. Additionally, in terms of flexion and extension regimes, symmetrical patterns were observed for all three screw fixation angle combinations, as expected. These characterizations further validate the effectiveness of the present phase field modeling as a powerful tool for capturing damage in biological tissues. Moreover, they demonstrate good agreement with recent works found in the literature \cite{molinari2021biomechanical,molinari2021effect}. 
	
	\begin{figure}[H]\centering
		\subcaptionbox{Back view of the evolution of the fracture pattern under vertical extension loading.\label{fig:evolutionextensionback}}
		{ \includegraphics[width=1.0\linewidth]{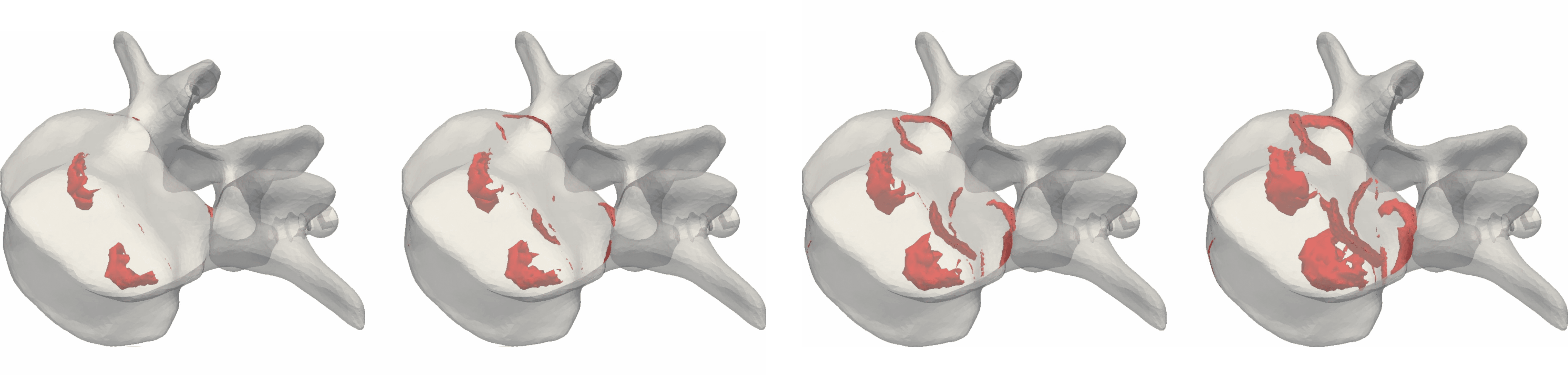}}
		\subcaptionbox{Back view of the evolution of the fracture pattern under vertical flexion loading regime. \label{fig:evolutionflexionback}}
		{\includegraphics[width=1.0\linewidth]{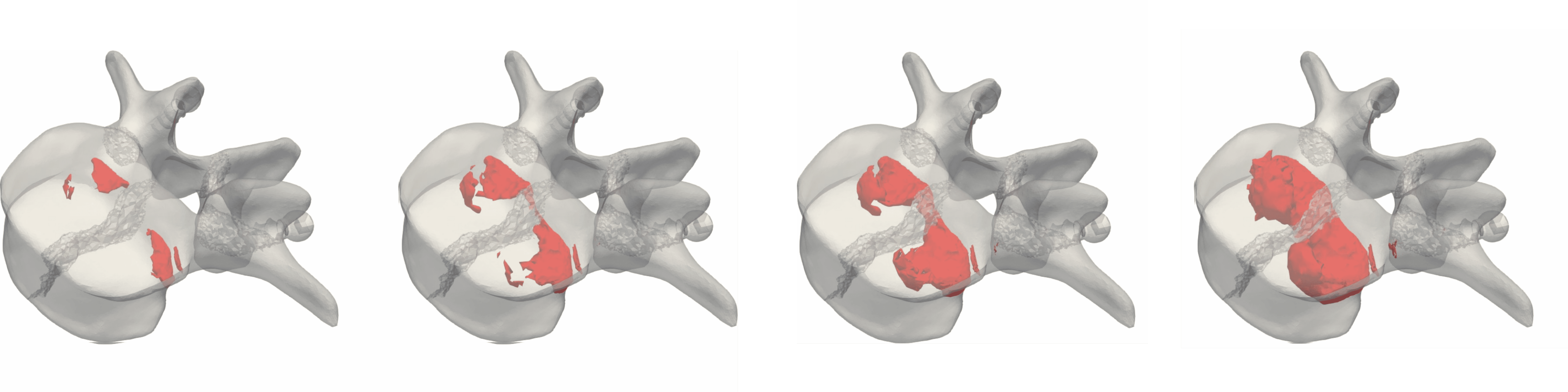}}
		\subcaptionbox{Back view of the evolution of the fracture pattern under a counter-clockwise axial rotation loading regime applied on the head of the screws.\label{fig:evolutioncounter-clockwise2back}}
		{\includegraphics[width=1.0\linewidth]{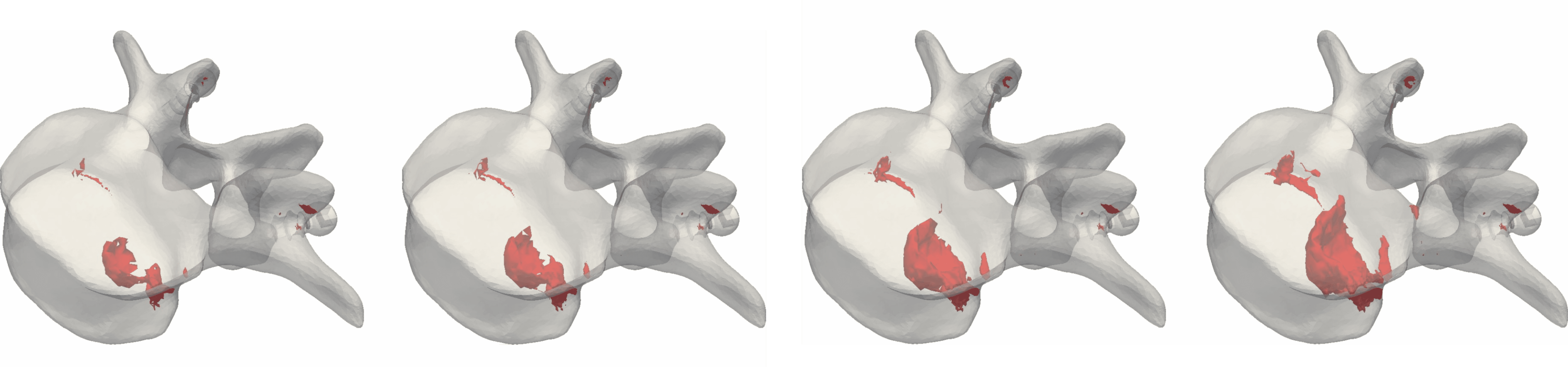}}
		\caption{Evolution of the damage pattern caused by loading regimes applied on the pedicle screws head: (a) extension motion; (b) flexion motion; and (c) counter-clockwise axial rotation motion.}
		\label{fig:evolutionspattern}
	\end{figure}
	
	\subsection{Staggered vs monolithic numerical schemes}
	\label{monovsstaggparagone}
	
	A numerical comparison has been conducted to assess the performances of both the staggered and monolithic numerical schemes, as described in Secs. \ref{sec:staggered} and \ref{sec:monolithic}, respectively, in terms of their accuracy in reproducing the fracture patterns and the Load vs. Displacement curves.
	
\begin{figure}[H]\centering
		{ \includegraphics[width=.7\linewidth]{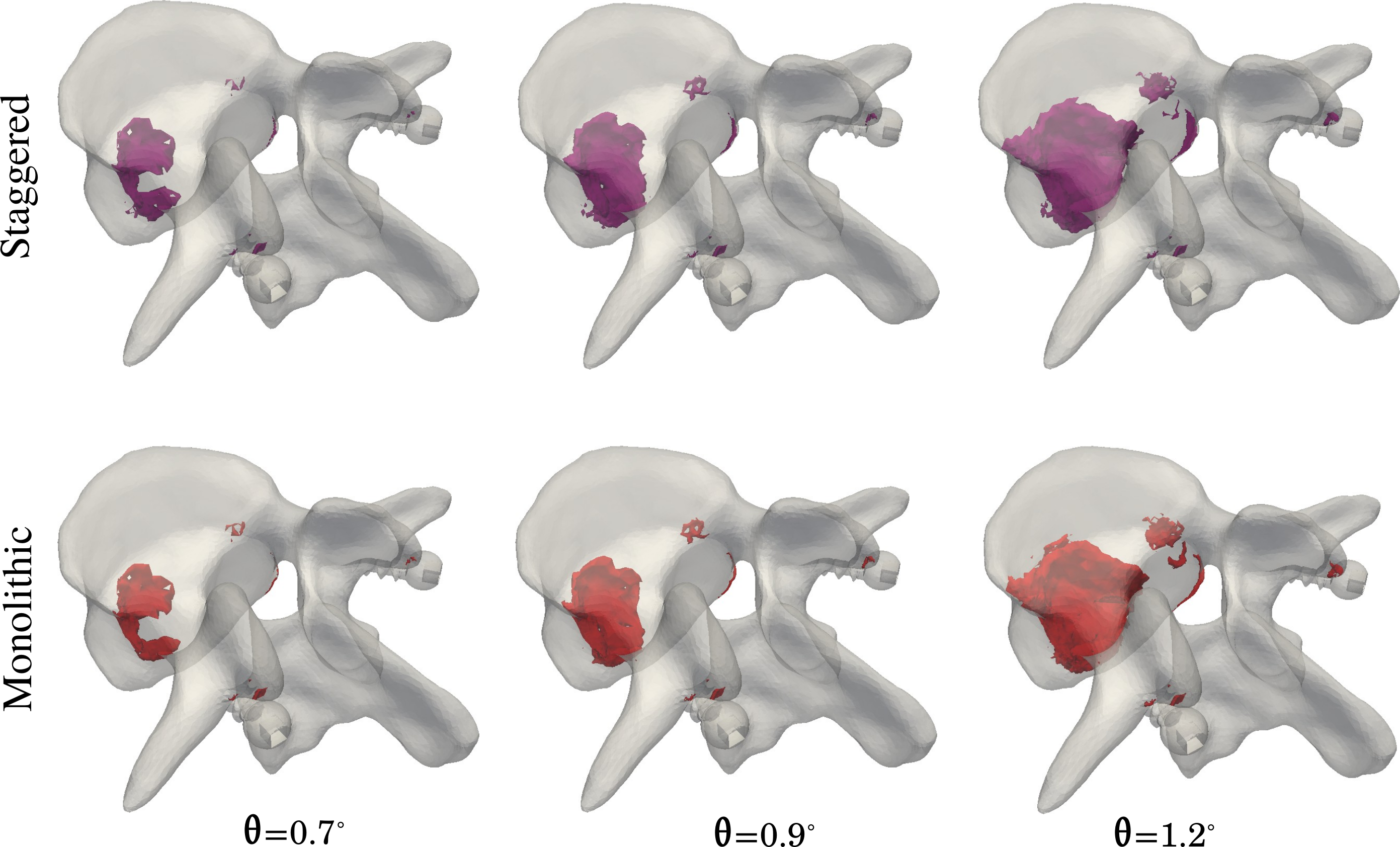}}
		\caption{Damage pattern comparisons between staggered and monolithic phase field models on a counter-clockwise axial rotation regime. }
		\label{fig:COMPARISON-STAGG-MONOL}
	\end{figure}

	For this comparative analysis, the discretized vertebra-screws model consisting of 300.000 elements has considered, with the screws inserted at an angle combination of ${\vec{\alpha}}=(-5,0),$ in a counter-clockwise torsion motion. It is notable that the fracture pattern obtained from the monolithic method bears resemblance to the damage patterns observed in the staggered simulation, see Fig. \ref{fig:COMPARISON-STAGG-MONOL}. Furthermore, the fractioned volume outcomes are similar for both approaches. It can also be observed that, in terms of accuracy, the Load vs Displacement curves are more accurate in the monolithic scheme due to the fact that the binding reaction on the vertebra endplate is zero and deviates very little. Nevertheless, considering the computational time consumed, the advantages of using the staggered scheme outweigh the potential benefits offered by the monolithic technique. While the staggered simulation took approximately $3.5$ hours to complete, the monolithic running time exceeded $15$ hours.	
	
	\begin{figure}[H]\centering
		\subcaptionbox{Applied loading on a counter-clockwise axial rotation loading regime applied on the head of the screws as function of the loading step.\label{fig:volfractured}}
		{ \includegraphics[width=.45\linewidth]{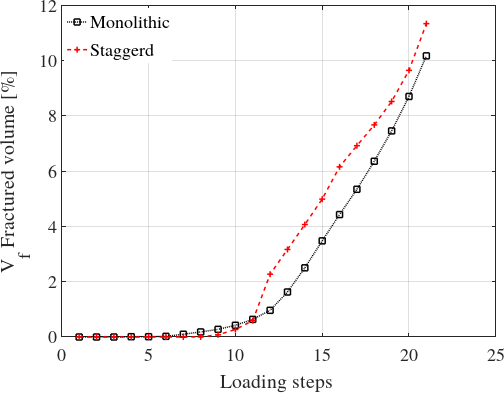}}
		\subcaptionbox{Fractured volume of the vertebra on a counter-clockwise axial rotation loading regime applied on the head of the screws as the loading step increases.\label{fig:FORCEvsLOADINGSTEPS}}
		{ \includegraphics[width=.45\linewidth]{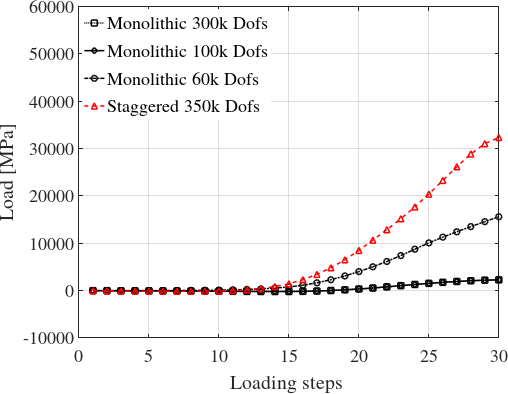}}
		\caption{Staggered and monolithic phase field models contrasted on a counter-clockwise axial rotation regime.}
		\label{fig:Constrated_mono_stagge}
	\end{figure}
	
	
	\section{Conclusions}
	
	A careful assessment of the fracture patterns and associated mechanisms in a human vertebra after the insertion of pedicle screws was scrutinized by implementing a finite element phase field method. Throughout the study, several aspects were pondered to accurately investigate the fractures and crack trajectories, such as maximum yield stress, critical energy release rate, and the characteristic length scale.
	
	The analysis of various fracture types was ensured by conducting a mesh sensitivity study, which provided an optimal mesh size balancing simulation running time and the model's capability to reproduce outcomes within a small relative error. These findings were supported by comparing Force vs. Displacement curves for all considered mesh refinements at the extension vertebral motion mode and a screws insertion angle configuration of $\vec{\alpha}=(-5,0).$ It is worth mentioning that the damage responses in the vertebra from the developed phase field model are affected when different configurations of the pedicle screws fixation angle are simulated.
	
	In addition, a comparison between phase field finite element approaches was also performed. Essentially, the staggered phase field model makes headway in terms of optimizing computational time consumption in relation to characterizing the damage within the vertebra-screws model, when compared to the results obtained from the monolithic phase field scheme.

	\addcontentsline{toc}{chapter}{References}
	\bibliographystyle{unsrtnat}
	\bibliography{references}

\end{document}